\documentclass[twoside,floatfix]{revtex4}
\usepackage{amsmath,graphicx,amssymb,fancyhdr,amsthm}

\theoremstyle{definition}

\theoremstyle{remark}

\def\beq{\begin{eqnarray}}
\def\eeq{\end{eqnarray}}

\def\Ri2{R_{\mu\nu}R^{\mu\nu}}

\begin{document}

\title{{\LARGE Simple Types of Anisotropic Inflation}}
\author{{\large \textbf{John D. Barrow$^{1}$ and Sigbj{\o }rn Hervik$^{2}$} }
\vspace{0.3cm}}
\affiliation{$^1$DAMTP, Centre for Mathematical Sciences, \\
Cambridge University, \\
Wilberforce Rd., Cambridge CB3 0WA, UK \vspace{0.4cm}\\
$^{2}$Faculty of Science and Technology,\\
University of Stavanger,\\
N-4036 Stavanger, Norway \vspace{0.3cm} }
\email{J.D.Barrow@damtp.cam.ac.uk, sigbjorn.hervik@uis.no}
\date{\today}

\begin{abstract}
We display some simple cosmological solutions of gravity theories with
quadratic Ricci curvature terms added to the Einstein-Hilbert lagrangian
which exhibit anisotropic inflation. The Hubble expansion rates are constant
and unequal in three orthogonal directions. We describe the evolution of the
simplest of these homogeneous and anisotropic cosmological models from its
natural initial state and evaluate the deviations they will create from
statistical isotropy in the fluctuations produced during a period of
anisotropic inflation. The anisotropic inflation is not a late-time
attractor in these models but the rate of approach to a final isotropic de
Sitter state is slow and is conducive to the creation of observable
anisotropic statistical effects in the microwave background. The statistical
anisotropy would not be scale invariant and the level of statistical
anisotropy will grow with scale.
\end{abstract}

\maketitle

% \vspace{0.3cm} \\
%EndAName

\pagestyle{fancy} \fancyhead{} % clear all header fields
\fancyhead[EC]{J.D. Barrow and S. Hervik} \fancyhead[EL,OR]{\thepage} %
\fancyhead[OC]{Anisotropic Inflation} \fancyfoot{} 
% clear all footer fields

\section{Introduction}

\ The observations of the cosmic microwave background (CMB) made by WMAP and
ground-based detectors are in good general agreement with the expectations
of a post-inflationary universe containing a residual quintessence field 
\cite{Hinshaw, komatsu}. However, the data quality has led to a focus upon
the presence of several unexpected features of the sky maps as well as a
search for any evidence of non-gaussianity in their statistics \cite{kam}.
In particular, there have been studies of the significance and possible
explanations for a lower than expected power in the quadrupole signal \cite%
{Copi}, possible alignments between low multipoles giving rise to a
preferred direction, or 'axis', on the sky \cite{axis} that can arise in
some anisotropic universes \cite{bsonoda}, and an apparent asymmetry between
the northern and southern ecliptic hemispheres \cite{asym}. These deviations
from statistical isotropy in the data are hard to assess definitively by
means of \textit{a posteriori} statistics and may be due to unnoticed biases
in the foreground subtraction, but careful studies of this potential problem
have yet to find evidence of such an effect at a level which can explain the
observations \cite{Gold}. A detailed study of the evidence for homogeneous
statistical anisotropy in the CMB has been conducted by Hanson and Lewis 
\cite{han}, who use quadratic maximum-likelihood estimators to analyse
Gaussian models with statistical anisotropy and realistic sources of
instrumental noise. They find evidence for anisotropy in the power spectrum
with a large angular dependence for the quadrupole, aligned close to the
ecliptic plane. Since this is the plane in which the satellite moves there
is a suspicion that this observed asymmetry may be associated with a
systematic instrumental effect or a beam ellipticity which is not corrected
for in the CMB maps analysed.

There have been several attempts to explore the possibility that small
anisotropic features might be imprinted upon the primordial fluctuation
spectrum by the process of inflation. In the standard general relativistic
model of inflation driven by scalar fields which violate the strong-energy
condition during a period of slow rolling, this would only be possible in
the very earliest moments of inflation if the prior state was one of extreme
expansion- and curvature-anisotropy because the dynamics rapidly approach
those of the de Sitter metric in the presence of a positive effective
cosmological constant. This situation is studied in ref \cite{pitrou}, but
only for the simplest form of expansion anisotropy ignoring the effects of
collisionless particles \cite{jb1} and curvature anisotropies \cite{jb2,DKLR},
which both make the residual effects larger. It is also possible to induce
local expansion anisotropies by the presence of super-horizon scale
inhomogeneities, and this scenario is explored by Gao \cite{gao}. If we
change the underlying inflationary model by adding a vector field \cite%
{vec,ACW,WKS}, a Chern-Simons or Kalb-Ramond field \cite{kal}, or some quadratic
curvature corrections to the gravitational lagrangian \cite{li}, then the
situation can change. Effective stresses are created which (unlike the
scalar inflaton field in general relativity) can violate the dominant or the
weak-energy conditions as well as the strong-energy condition. Under these
circumstances the de Sitter metric may no longer be an attractor during
accelerated expansion and anisotropic inflationary behaviour might even be
possible. Anisotropic inflation is only possible in general relativity with
a positive cosmological constant and matter that obeys the strong energy
condition if the 3-curvature is positive. There is a known exact $%
S^{1}\times S^{2}$ Kantowski-Sachs universe of \ this type \cite{Weber}
which expands forever, with exponential expansion of its scale factors in
two directions and a constant scale factor along the third. This model has
been used in a study of inflation \cite{LZ}, however, its anisotropic
behaviour is unstable within this class of Kantowski-Sachs models and to the
addition of matter fields \cite{BY}.

In this paper we are interested in the situation where a quadratic Ricci
term, $R_{\mu\nu}R^{\mu\nu}$, is added to the Einstein-Hilbert term. We have already
shown \cite{BH2} that in the resulting theory with positive cosmological
constant it is possible to find simple exact spatially homogeneous
cosmological models of Bianchi type I which inflate anisotropically. Here we
explore these solutions in more detail and show that the same cosmological
models also possess an exact de Sitter attractor solution. However, the
evolution towards the de Sitter attractor from an earlier state of
anisotropic inflation is unusually slow and there is ample opportunity for
anisotropic statistical effects to be imprinted upon the fluctuation
spectrum created by the accelerated expansion. These anisotropies will be
larger at earlier times in the inflationary phase and therefore will imprint
greater anisotropic effects on larger scales than on smaller ones. This
model is mathematically simple, with Euclidean space sections, and provides
a useful testing ground for computing more detailed effects of anisotropic
inflation on the scalar and tensor irregularity spectra. It allows us to
determine the sense and relative magnitudes of explicit expansion
anisotropies during a period of volume inflation in which the orthogonal
scale factors expand by different exponential time factors.

Usually inflation is considered to be driven by an isotropic scalar field; here we will consider a simple cosmologial constant as an example of such an effect. The models can easily be extended to include isotropic scalar fields of this kind, however, this may require additional fine-tuning for the scalar field to come to dominate the expansion dynamics at the right time. This is also the case for $f(R)$ effects through the standard conformal transformation \cite{conf}. The effects from the quadratic Ricci term, on the other hand, are different as they allow for anisotropic degrees of freedom -- indeed, a quadratic Ricci term is the main source of the anisotropic inflation considered in this paper.

\section{A Bianchi type I model}

Our staring point is the equations of motion using a dynamical systems
approach. For the Bianchi type I (and II) models, these are given in \cite%
{BH2}. We will consider the quadratic theory where the Einstein-Hilbert
action is modified: 
\begin{equation*}
S_{G}=\frac{1}{2\kappa }\int_{M}\mathrm{d}^{4}x\sqrt{|g|}\left( R+\alpha
R^{2}+\beta \Ri2-2\Lambda \right) .
\end{equation*}%
Upon variation, the equations of motion are the modified Einstein equations: 
\begin{equation}
G_{\mu \nu }+\Phi _{\mu \nu }+\Lambda g_{\mu \nu }=0,  \label{field}
\end{equation}%
where $G_{\mu \nu }=R_{\mu \nu }-(1/2)Rg_{\mu \nu }$ is the regular Einstein
tensor, and 
\begin{eqnarray}
\Phi _{\mu \nu } &=&2\alpha R\left( R_{\mu \nu }-\frac{1}{4}Rg_{\mu \nu
}\right) +(2\alpha +\beta )\left( g_{\mu \nu }\Box -\nabla _{\mu }\nabla
_{\nu }\right) R  \notag \\
&&+\beta \Box \left( R_{\mu \nu }-\frac{1}{2}Rg_{\mu \nu }\right) +2\beta
\left( R_{\mu \sigma \nu \rho }-\frac{1}{4}g_{\mu \nu }R_{\sigma \rho
}\right) R^{\sigma \rho },
\end{eqnarray}%
and $\Box \equiv \nabla ^{\mu }\nabla _{\mu }$. We note that the GR-limit
can be obtained by letting $(\alpha ,\beta )\rightarrow (0,0)$. Now consider
the spatially homogeneous Bianchi type I metrics. We can always write their
metric line-elements as 
\begin{equation*}
\mathrm{d}s^{2}=-\mathrm{d}t^{2}+\delta _{ab}{\mbox{\boldmath${\omega}$}}^{a}%
{\mbox{\boldmath${\omega}$}}^{b},
\end{equation*}%
where ${\mbox{\boldmath${\omega}$}}^{a}$ is a triad of one-forms which, for
the Bianchi type I model, can be written as: ${\mbox{\boldmath${\omega}$}}%
^{a}=\mathsf{e}_{~i}^{a}(t)\mathrm{d}x^{i}$.

Defining the time-like hypersurface-orthogonal vector $\mathbf{u}=\partial
/\partial t,$ we can define the Hubble scalar, $H$, and the shear, $\sigma
_{ab}$, as follows: 
\begin{equation*}
H\equiv \frac{1}{3}u_{;\mu }^{\mu },\quad \sigma _{ab}=u_{(a;b)}-H\delta
_{ab}.
\end{equation*}%
We will also restrict attention to cosmological models where the shear is
diagonal, so we can write 
\begin{equation*}
\sigma _{ab}=\text{diag}(-2\sigma _{+},\sigma _{+}+\sqrt{3}\sigma
_{-},\sigma _{+}-\sqrt{3}\sigma _{-}).
\end{equation*}

We define the dimensionless expansion-normalised variables by scaling out
appropriate powers of $H$%
\begin{eqnarray}
&&B=\frac{1}{(3\alpha +\beta )H^{2}},\quad \chi =\frac{\beta }{3\alpha
+\beta },  \notag \\
&&Q=\frac{\dot{H}}{H^{2}},\quad Q_{2}=\frac{\ddot{H}}{H^{3}},\quad \Omega
_{\Lambda }=\frac{\Lambda }{3H^{2}},  \notag \\
&&\Sigma _{\pm }=\frac{\sigma _{\pm }}{H},\quad \Sigma _{\pm 1}=\frac{\dot{%
\sigma}_{\pm }}{H^{2}},\quad \Sigma _{\pm 2}=\frac{\ddot{\sigma}_{\pm }}{%
H^{3}}.
\end{eqnarray}%
Note the presence of time derivatives of the variables $Q_{2}$ and $\Sigma
_{\pm 2};$ this reflects the $4^{th}$-order time derivatives in the field
equations of the quadratic theory\footnote{%
The variable $Q_{2}$ can be related to the statefinders $q$ and $j$ by $%
Q_{2}=j+3q+2$.}; $\chi $ is a constant. We also introduce the dynamical time
variable $\tau $ by 
\begin{equation*}
\frac{\mathrm{d}\tau }{\mathrm{d}t}=H,
\end{equation*}%
and note that since $H=\dot{a}/a$, the dynamical time is related to the
length scale as $a=\ell_0e^\tau$, where $\ell_0>0$ is a constant. We also assume that the cosmological
constant is positive: $\Omega _{\Lambda }>0$.

The full set of dynamical equations for the Bianchi type I model (and type
II model) are given in \cite{BH2}. In \cite{BH2} it was shown that there
exists a peculiar set of solutions for this theory, namely a set of
anisotropically inflating solutions. A stability analysis of these solutions
was also performed. Here, we intend to study these solutions further; in
particular, we will investigate their possible influence on any statistical
relic anisotropy at the end of inflation.

For later reference, let us also give the equations of motion. We will, for
simplicity, consider the invariant subspace when $(\Sigma _{+},\Sigma
_{-})\propto (\Sigma _{+1},\Sigma _{-1})\propto (\Sigma _{+2},\Sigma _{-2})$%
. This enables us to write: 
\begin{eqnarray}
&&(\Sigma _{+},\Sigma _{-})=\Sigma (\cos \phi ,\sin \phi ),\quad (\Sigma
_{+1},\Sigma _{-1})=\Sigma _{1}(\cos \phi ,\sin \phi ),  \notag \\
&&(\Sigma _{+2},\Sigma _{-2})=\Sigma _{2}(\cos \phi ,\sin \phi ),\quad \phi
^{\prime }=0.
\end{eqnarray}%
The equations of motion are now: 
\begin{eqnarray}
B^{\prime } &=&-2QB, \\
\Omega _{\Lambda }^{\prime } &=&-2Q\Omega _{\Lambda }, \\
Q^{\prime } &=&-2Q^{2}+Q_{2}, \\
Q_{2}^{\prime } &=&-3(Q+2)Q_{2}-\frac{9}{2}(Q+2)Q-\frac{3}{4}B\left(
1+\Sigma ^{2}-\Omega _{\Lambda }+\frac{2}{3}Q\right)   \notag \\
&&-\frac{3}{2}(1+2\chi )\Sigma ^{4}-\frac{1}{4}(8+\chi )\Sigma
_{1}^{2}-(4-\chi )\Sigma \Sigma _{1}  \notag \\
&&-\frac{1}{4}(4-\chi )(3\Sigma ^{2}+2\Sigma \Sigma _{2}+2Q\Sigma ^{2}), \\
\Sigma ^{\prime } &=&-Q\Sigma +\Sigma _{1}, \\
\Sigma _{1}^{\prime } &=&-2Q\Sigma _{1}+\Sigma _{2}, \\
\Sigma _{2}^{\prime } &=&-3(Q+2)\Sigma _{2}+\frac{\Sigma _{1}}{\chi }\left[
B-(11\chi -8)+4Q(1-\chi )+4\Sigma ^{2}(1+2\chi )\right]   \notag \\
&&+\frac{\Sigma }{\chi }\left[ 3B+(4-\chi )(6+Q_{2}+7Q)+4(1+2\chi )(3\Sigma
^{2}+2\Sigma \Sigma _{1})\right]   \label{eqSigma+2}
\end{eqnarray}%
These equations are subject to the ('Friedmann-like') constraint: 
\begin{eqnarray}
0 &=&B(1-\Omega _{\Lambda }-\Sigma ^{2})+12Q-2Q^{2}+4Q_{2}-(4-\chi
)(3+2Q)\Sigma ^{2}  \notag \\
&&-6(1+2\chi )\Sigma ^{4}-\chi (\Sigma _{1}^{2}-2\Sigma \Sigma
_{2})+4(2+\chi )\Sigma \Sigma _{1}.
\end{eqnarray}

The parameter $Q$ is related to the usual deceleration parameter $q$ via 
\begin{equation*}
q=-(1+Q),
\end{equation*}%
and the variable $B$ measures how greatly the quadratic part of the
lagrangian dominates over the general-relativistic Einstein-Hilbert term $%
R-2\Lambda $. In particular, the larger the value of $B$, the "closer" we
are to GR. The $B=0$ case corresponds to a purely quadratic lagrangian
theory whose equations of motion reduce to $\Phi _{\mu \nu }=0$.

\subsection{The Inflating solutions}

We will now focus on two (sets of) solutions, namely, the de Sitter
solution, and the anisotropically inflating type I solutions.

\paragraph{The de Sitter solution, $\mathrm{dS}$:}

The de Sitter solution is characterised by the critical points where 
\begin{equation*}
Q=Q_{2}=\Sigma _{\pm }=\Sigma _{\pm 1}=\Sigma _{\pm 2}=N=0,\quad \Omega
_{\Lambda }=1,\quad B\neq 0.
\end{equation*}%
Its stability is assured if \cite{BH2} 
\begin{equation*}
B>0\Rightarrow (3\alpha +\beta )>0,\quad \frac{B+2(4-\chi )}{\chi }%
<0\Rightarrow \frac{1+2\Lambda (4\alpha +\beta )}{\beta }<0.
\end{equation*}

\paragraph{Anisotropically-inflating type I universes, $\mathcal{A}(I)$:}

For certain values of $\chi $ and $B$, there are also exact solutions that
describe anisotropic inflationary solutions of Bianchi type I \cite{BH2}: 
\begin{eqnarray}
&&(\Sigma _{+},\Sigma _{-})=\Sigma (\cos \phi ,\sin \phi ),\quad \Sigma
^{2}=\Sigma _{0}^{2}\equiv -\frac{2(4-\chi )+B}{4(2\chi +1)},  \notag \\
&&Q=\Sigma _{\pm 1}=\Sigma _{\pm 2}=N=0.  \notag
\end{eqnarray}%
There are two classes of such solutions, depending on the values of $B$ and $%
\Omega _{\Lambda }$. Here we will concentrate on the following: 
\begin{equation*}
B=\mathrm{constant},\quad \Omega _{\Lambda }=\frac{18\chi -B}{8(2\chi +1)}.
\end{equation*}%
So long as $\chi $ and $B$ take values for which $\Sigma _{0}^{2}>0,$ these
solutions exist. Moreover, the solution is unstable, but for $B>0$ contains
only \emph{one} unstable mode.

The metrics corresponding to the case where $B\neq 0$ can be written 
\begin{eqnarray}
&&\mathrm{d}s^{2}=-\mathrm{d}t^{2}+e^{2bt}\left[ e^{-4\sigma _{+}t}\mathrm{d}%
x^{2}+e^{2(\sigma _{+}+\sqrt{3}\sigma _{-})t}\mathrm{d}y^{2}+e^{2(\sigma
_{+}-\sqrt{3}\sigma _{-})t}\mathrm{d}z^{2}\right] , \\
&&b^{2}=\frac{1+8\Lambda (\alpha +\beta )}{9\beta },\quad (\sigma
_{+}^{2}+\sigma _{-}^{2})=-\frac{1+2\Lambda (4\alpha +\beta )}{18\beta }. 
\notag
\end{eqnarray}%
We note that for these solutions to exist we need both these squares to be
positive. Furthermore, we require $\sigma _{-}^{2}+\sigma
_{+}^{2}+b^{2}/2=\Lambda /3>0$.

\section{Anisotropic inflation}

Based on the solutions above and their stability, \emph{there are values of
the parameters where $\mathcal{A}(I)$ exists, and the de Sitter solution is
stable.} In fact, the connection is deeper than this. Assuming $B>0$, and $%
\Lambda >0$, then if $\chi >4$, $\mathcal{A}(I)$ is connected to $\mathrm{dS}
$ via a bifurcation at $B=2(\chi -4)$. For $B<2(\chi -4)$ the de Sitter
solution is unstable (1 unstable mode). As the value of $B$ increases, $%
\mathrm{dS}$ bifurcates at $B=2(\chi -4)$ creating the equilibrium points $%
\mathcal{A}(I)$. For $B>2(\chi -4)$, $\mathcal{A}(I)$ has acquired the
unstable mode and made $\mathrm{dS}$ stable. Hence, the only unstable mode
of $\mathcal{A}(I)$ is actually connected to the de Sitter solution and
these inflationary solutions should therefore be connected via a
heteroclinic orbit (at least close to the bifurcation point). It is this
fact we shall exploit here because this opens up the possibility for that the
universe approaches the point $\mathcal{A}(I)$ and so starts to inflate
anisotropically. This state is almost stable but it does have an unstable
mode. This unstable mode will therefore eventually drive the evolution away
from anisotropic inflation, towards isotropic inflation (represented by the
stable point $\mathrm{dS}$).

\begin{figure}[tbp]
\centering \includegraphics[width=10cm,clip=true]{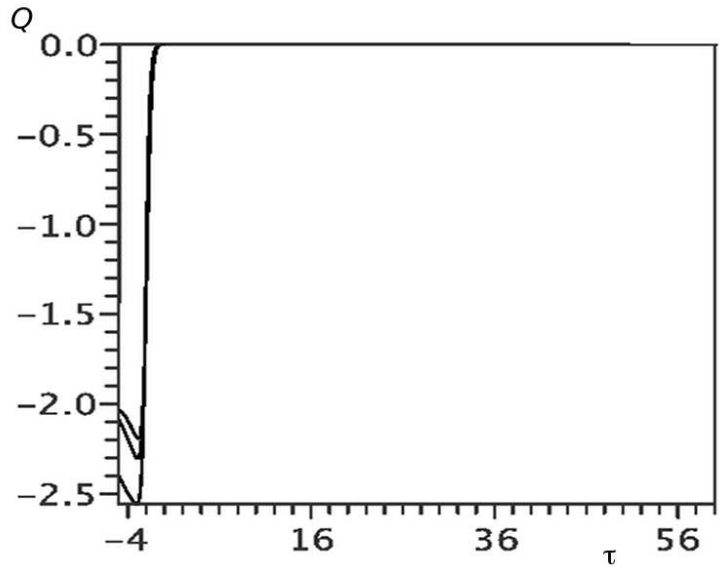}\newline
\includegraphics[width=10cm,clip=true]{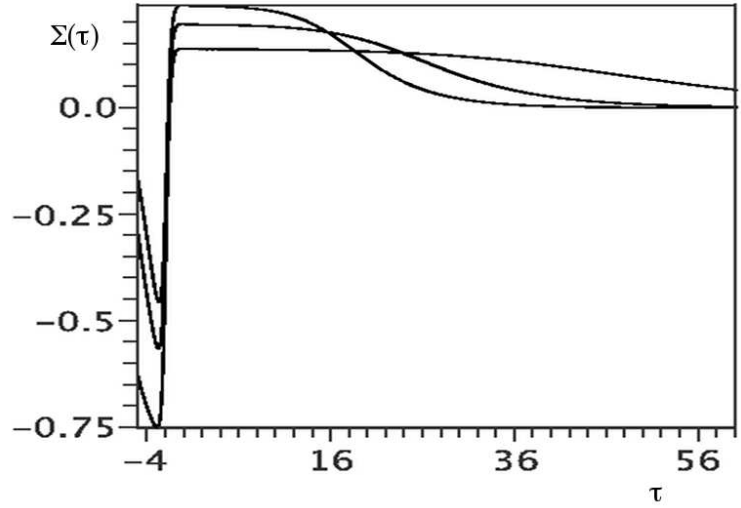}
\caption{Inflation: The evolution of $Q$ and $\Sigma $ with time, $\protect%
\tau $, for universes with three different initial values. Here, we set $%
\protect\chi =6$.}
\label{fig1}
\end{figure}

In Figure \ref{fig1} the evolution of universes with three different initial
values are plotted. The universes start to inflate around $\tau =0$, for
which $Q\approx 0$, however, as can be seen, the shear is non-zero; hence,
the universe starts inflating anisotropically. During a transient period the
universe can stay arbitrary close to the anisotropically inflating point $%
\mathcal{A}(I)$; however, as we can see, eventually the universe will move
towards isotropic inflation. For the universes displayed, we see that during
the time from $\tau =0$ to $\tau =60$ (which corresponds to 60 e-folds), we
have $Q\approx 0$ and therefore they are inflating during the entire period,
even if the universes are undergoing a transition between anisotropic
inflation and isotropic inflation.

Interestingly, the smaller the shear is at $\tau =0$, the longer it takes to
enter the state of isotropic inflation. The reason for this can be seen from
the unstable mode of $\mathcal{A}(I)$. From the eigenvalues derived in \cite%
{BH2}, we see that the unstable mode goes as $\propto \exp ({\lambda
_{1}\tau })$, where $\lambda _{1}=(3/2)(\sqrt{1+8\Sigma _{0}^{2}}-1)$.
Consequently, the smaller the shear, the more time it would take for the
universe to enter a state of isotropic inflation. Furthermore, since the
universe inflates during the transition, $Q\approx 0$ and the value of $%
B\approx constant$. Therefore, close to the de Sitter isotropic point the
shear will decay as $\Sigma \propto \exp (\lambda _{2}\tau )$ where $\lambda
_{2}$ can be estimated to be 
\begin{equation*}
\lambda _{2}\approx -\frac{3}{2}\left( 1-\sqrt{1-\frac{16(2\chi +1)}{9\chi }%
\Sigma _{0}^{2}}\right) .
\end{equation*}%
For small $\Sigma _{0},$ this reduces to $\lambda _{2}\approx -4(2\chi
+1)\Sigma _{0}^{2}/(3\chi )$. This implies that the shear during the
anisotropic inflation, $\Sigma _{0}$, is imprinted in the decay rate of the
shear on the approach to isotropic de Sitter state and the smaller the shear 
$\Sigma _{0}$, the slower the approach towards the isotropic de Sitter
inflation.

A consequence of this is that at the end of 60 $e$-folds ($\tau=60$), the
universe may still have considerable amount of shear in spite of the fact
that the universe has inflated. However, as we can see from Figure \ref{fig1}%
, this shear can also be arbitrary small.

\begin{figure}[tbp]
\centering \includegraphics[width=10cm,clip=true]{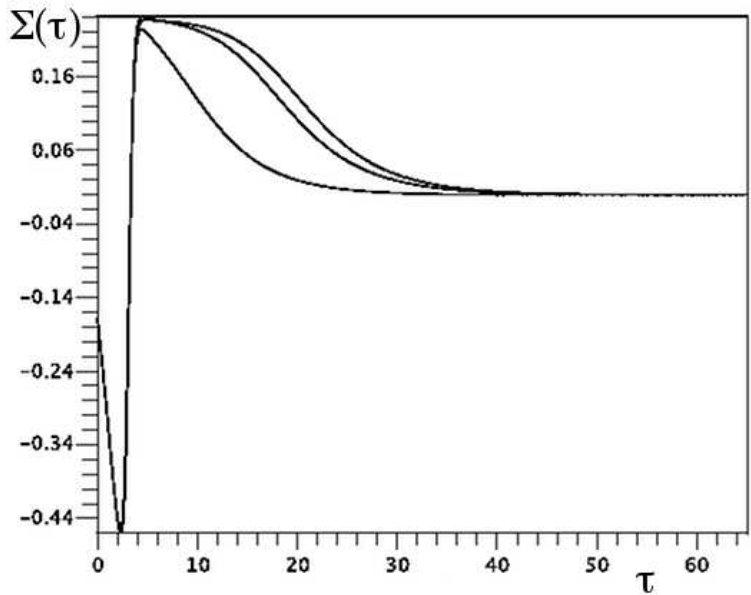}
\caption{The evolution of the dimensionless expansion-normalised shear, $%
\Sigma $, with time, $\protect\tau $, for three universes starting away from
the inflating solutions. They experience a transient period during which the
universes inflate anisotropically. The evolution displayed requires some
fine-tuning but the plot illustrates that there is a set of non-zero measure
that gives the desired behaviour.}
\label{fig2}
\end{figure}

In Figure \ref{fig2} the numerics have been started away from the
anisotropic inflationary solutions. With a bit of fine tuning we can see
that the solutions experience a transient period during which the universe
inflates anisotropically. This shows that there is a set of non-zero measure
of initial values that experience anisotropic inflation.

\subsection{Consequences for the CMB}

In ref. \cite{ACW}, the imprints of a preferred direction on the CMB was
studied for an anisotropic model in which rotational invariance is broken by
the presence of a vector picking out a preferred direction with unit vector $%
\hat{n}$. If parity is preserved ($\vec{k}\rightarrow -\vec{k}$) then the
leading order of the anisotropic power spectrum has a quadrupole form, with

\begin{equation*}
P(\vec{k})=P(k)[1+g(k)(\hat{n}\cdot \hat{k})^{2}+higher\text{ }order],
\end{equation*}%
where $P(k)$ is the isotropic part of the power spectrum and $g(k)$ measures
the power of the statistical anisotropy. If the anisotropic contributions
are scale independent then $g(k)$ will be a constant. However, this need not
always be the case, as in the study of anisotropies induced by super-horizon
inhomogeneities \cite{gao}.

The model considered in ref \cite{ACW} was an axisymmetric anisotropically
inflationary model. Our model allows for axisymmetry in the special case $%
\Sigma _{-}=0$ (the general case has no rotational symmetry whatsoever).
Their model actually corresponds to the exact equilibrium point, but in our
model this is unstable and evolves toward the isotropic de Sitter state. In
their eq.(39) they define the anisotropy parameter $\epsilon _{H}$, which in
our notation is just the dimensionless shear distortion (in the special case 
$\Sigma _{-}=0$): 
\begin{equation*}
g(k)\propto \epsilon _{H}\equiv \frac{2\sigma _{+}}{H}=2\Sigma ,
\end{equation*}%
(recall that the dynamical time $\tau $ is related to the scale factor $a$
through $a=a_{0}e^{\tau }$). In our model $g(k)$ will evolve and so cause
the perturbation spectrum to depend on the time when each perturbation scale
left the horizon. Hence, the statistical anisotropy will not be scale
invariant. In our models the anisotropy is larger at the beginning of the
inflation which means that long wavelengths will be more anisotropic than
the short wavelengths.

\subsection{Other Bianchi models}

The model considered here is a very special Bianchi model. However, the
existence of anisotropic inflation seems to be a widespread feature of
quadratic theories of gravity. In \cite{BH1} we showed the existence of
anisotropic inflation for Bianchi type II models. Among the most general
Bianchi models are the Bianchi type VI$_{h}$ and VII$_{h}$ models and
anisotropic inflation is also present there. The following metrics can be
shown to be solutions to $\Phi _{\mu \nu }=0$ in eq. (\ref{field}): 
\begin{equation*}
\mathrm{d}s^{2}=-\mathrm{d}t^{2}+\mathrm{d}x^{2}+e^{2r(t+x)}\left[ e^{2a}(A%
\mathrm{d}y+B\mathrm{d}z)^{2}+e^{-2a}(C\mathrm{d}y+A\mathrm{d}z)^{2}\right] .
\end{equation*}%
Here, $r$ and $a$ are constants. For the various Bianchi types, the
functions $A$, $B$ and $C$ are as follows:

\begin{enumerate}
\item {} Type VII$_h$: $A=\cos[\omega(x+ t)]$, $B=-C= \sin[\omega(x+t)]$.
\item {} Type VI$_h$: $A=\cosh[\omega(x+ t)]$, $B=C= \sinh[\omega(x+t)]$.
\item {} Type IV: $A=1$, $B= \omega(x+t)$, $C=0$.
\end{enumerate}

There are no analogues of these solutions in general relativity because they
would violate the weak energy condition. Interestingly, these are also
so-called plane-wave spacetimes. We note that the 3-space volume expands
exponentially, $V\propto \exp (2rt)$ in terms of the comoving proper time, $t
$; hence, these are indeed inflationary solutions. However, as we see that
one orthogonal direction is actually fixed and does not expand at all.

The existence of such solutions for these Bianchi types indicates that
anisotropic inflation is a more general feature of quadratic theories than
previous thought and is present in even the most general class of
anisotropic universes, which includes types $VI_{h}$ and $VII_{h}$.

\section{Conclusions}

Simple Bianchi type I universes in gravity theories in which the
Einstein-Hilbert lagrangian is augmented by the addition of terms quadratic
in the scalar curvature ($R^{2}$) and Ricci invariant ($R_{\mu\nu}R^{\mu\nu}$)
display evolution that commences from a near-isotropic singularity. It
mimics the behaviour of a radiation-dominated Friedmann universe \cite{mid}
and asymptotes towards a de Sitter late-time attractor. However, the
evolution spends a long time evolving slowly through a period of anisotropic
inflation during which the 3-volume expands exponentially in comoving proper
time and the three directional scale factors increase at different rates. We
display exact solutions which display this transitional anisotropic
inflationary behaviour and show that it does not arise in the general
relativity limit when the higher-order curvature terms vanish from the
lagrangian. With a small amount of fine tuning of the pre-inflationary
evolution this behaviour can leave a distinctive imprint on the spectrum of
inhomogeneities created by any period of inflation defined by the
exponential increase in the 3-volume. Using the characterisation of the
leading statistically anisotropic contribution to the power spectrum of
inhomogeneities that was introduced by Ackerman, Carroll and Wise \cite{ACW}%
, we determine the amplitude of the statistical anisotropy in the spectrum
and show that it will not be scale invariant because the anisotropic effects
increase with scale.

There have been a number of studies of the possible sources of statistical
anisotropy in the power spectrum of the microwave background. The most
detailed study \cite{han} finds significant evidence for anisotropy but no
proof that it is primordial in origin. Our analysis identifies a broad class
of higher-order gravity theories in which there are solutions that differ
significantly from those in general relativity. In particular,
ever-expanding vacuum solutions with a positive cosmological constant do not
approach the de Sitter solution. The higher-order curvature terms contribute
effective stress terms which violate the energy conditions that are needed
for the cosmic no hair theorems of general relativistic cosmology to hold.
Hence, they allow new types of exact solution to exist in which different
directions accelerate at different rates. We have shown that solutions of
this anisotropic inflationary type also exist in some of the most general
Bianchi type universe and are not confined to the type I case we have used
for simplicity. In the general Bianchi I universe we have studied, there is
attraction to an asymptotic de Sitter solution, as in general relativity,
but the evolution spends a large number of e-folds of expansion in the
neighbourhood of an anisotropic inflationary solution. The existence of such
behaviour near the Planck scale and the persistence of anomalies in the sky
maps derived so far from observations of the microwave sky suggest that
there is a possibility the two are related.

\end{document}